\documentclass[letterpaper,10pt]{article}
\usepackage{setspace}
\usepackage{graphicx}
\usepackage{amsmath}
\usepackage{array}
\usepackage{natbib} 
\usepackage{verbatim} 	
\usepackage[letterpaper,vmargin=1in,hmargin=1in]{geometry}

\newcommand{\fignameSP}{Fig.~}
\newcommand{\tabnameSP}{Table }

\begin{document}

\title{Progressive Collapse Mechanisms of Brittle and Ductile Framed Structures}
\maketitle

\begin{center}
\author{E. Masoero\footnotemark[1], 
F. K. Wittel\footnotemark[2], 
H. J. Herrmann\footnotemark[3], 
B. M. Chiaia\footnotemark[4].}
\end{center}

\footnotetext[1]{Ph.D. Stud., Politecnico di Torino, Department of Structural and Geotechnical Engineering, Corso Duca degli Abruzzi 24, 10129 Torino, Italy. Email address: enrico.masoero@polito.it}
\footnotetext[2]{Dr., ETH Zurich, Institute for Building Materials, Schafmattstrasse 6, 8093 Zurich, Switzerland. Email address: fwittel@ethz.ch}
\footnotetext[3]{Prof., ETH Zurich, Institute for Building Materials, Schafmattstrasse 6, 8093 Zurich, Switzerland. Email address: hans@ifb.baug.ethz.ch}
\footnotetext[4]{Prof., Politecnico di Torino, Department of Structural and Geotechnical Engineering, Corso Duca degli Abruzzi 24, 10129 Torino, Italy. Email address: bernardino.chiaia@polito.it}

\begin{abstract}
In this paper, we study the progressive collapse of 3D framed structures made of reinforced concrete after the sudden loss of a column. The structures are represented by elasto-plastic Euler Bernoulli beams with elongation-rotation failure threshold. We performed simulations using the Discrete Element Method considering inelastic collisions between the structural elements. The results show what collapse initiation and impact-driven propagation mechanisms are activated in structures with different geometric and mechanical features. Namely, we investigate the influence of the cross sectional size and reinforcement $\alpha$ and of the plastic capacity $\beta$ of the structural elements. We also study the final collapse extent and the fragment size distribution and their relation to $\alpha$, $\beta$ and to the observed collapse mechanisms. Finally, we compare the damage response of structures with symmetric and asymmetric reinforcement in the beams.
\end{abstract}

\section{Introduction}\label{Intro}
Local damage of buildings can either be due to accidental events like gas explosions, gross design-construction errors and malicious terrorist attacks \citep{Levy_Salvadori-1992}, or can be thoroughly planned as part of controlled demolition processes with blast. Subsequent cascades of failures can cause large economic and human loss when triggered by accidental events and make the difference between effective and dangerously ineffective demolitions.

Research about progressive collapse of buildings proceeded discontinuously since 1970s mostly prompted by outstanding and shocking catastrophes. Interest in the subject rose after the Ronan Point partial collapse in 1968 due to gas explosion. During the seventies, the fundamental approaches to structural robustness as well as many indicators, like the Reserve Strength Ratio (RSR) \citep{Maes_Fritzsons-2006}, were formulated, also with regard to off-shore structures that suffered brittle collapses on the North Sea. Renewed attention to the problem was given due to terrorist attacks against the Alfred P. Murrah federal building, Oklahoma City 1995, and the World Trade Center (WTC), New York 2001 \citep{Val_Val-2006}. 

Nowadays many codes prescribe alternate paths for the load (Alternate Load Path Method (ALPM) \citep{Val_Val-2006}) and high toughness of structural members and their interconnections \citep{Nair-2004}. Nonetheless, these measures are not always sufficient to prevent progressive collapse \citep{Vlassis_Nethercot-2006}. Moreover, even though the serious damage amplification due to dynamics has been exhaustively pointed out in \citep{Pretlove_Ramsden-1991,Marjanishvili_Agnew-2006}, static analyses are still used in the context of the ALPM.

Developing efficient tools to evaluate structural robustness and to prove the effectiveness of measures aimed at preventing progressive collapse is therefore an important issue and today several algorithms and models are available in literature. A simplified approach to take dynamics and impacts between falling elements into account was proposed by \citep{Vlassis-PHD-2007} 
and \citep{Chiaia_Masoero-2008} showed its analogy with the variational approach to fracture mechanics. The scheme is based on energy balance, requires only static analyses, and was effectively applied to strain hardening \citep{Vlassis_Nethercot-2006} and softening \citep{Chiaia_Masoero-2008} structural elements. 

Analytical 1D models were developed after the WTC collapse \citep{Bazant_Verdure-2007,Cherepanov_Esparragoza-2007,Seffen-2008}. In these models, progressive buckling of the columns is due to the impact of the upper floors, considered as an increasing falling mass. Differently, computer simulations permit to study more complex 2D and 3D structures. Several key factors that influence robustness of frames have already been identified. For instance, we know that the loss of external columns from the facades or the corners of a buildings are the most serious scenarios where, according to the ALPM method, one column is instantaneously removed (see, e.g. \citep{Kaewkulchai_Williamson-2003}). Moreover, it was shown that beam-column connections are critical points of failure initiation \citep{Khandelwal_El-Tawil-2008} and that catenary effects in the floor slabs can remarkably improve robustness \citep{Vlassis_Nethercot-2006}. 

Even though the final outcome of progressive collapse depends on the collisions between structural elements, most of literature focuses on collapse initiation. Collisions are rarely taken into account either detailed with Finite Elements \citep{Hartmann_Breidt-press,Luccioni_Ambrosini-2004}, or approximated in the framework of Finite Macro-Elements \citep{Kaewkulchai_Williamson-2006,Grierson_Xu-2005}. Detailed Finite Elements are too demanding in terms of computational time for extensive parametric studies on large structures. Differently, Finite Macro-Elements are efficient and can be applied to large structures, but they require strong approximations to take into account collisions and catenary effects, especially in 3D (e.g. see \citep{Isobe_Tsuda-2003}).

The lack of experimental results of progressive collapses suggests an approached based on simulations  whose reliability arises from the basic physics incorporated. The results obtained with such algorithms can be used to construct, test and calibrate simpler models. In this work, we use spherical Discrete Elements (DE) to simulate the progressive collapse of typical 3D framed structures made of reinforced concrete with fixed regular overall geometry (see Sec.~\ref{Simul}). The aim is to study the collapse initiation mechanisms due to dynamic stress redistribution, and the subsequent damage propagation mechanisms due to collisions between the structural elements. Understanding the activated mechanisms, depending on the strength, the stiffness and the plastic properties of the structural elements, can help to choose optimal robustness oriented design solutions as well as the most appropriate structural reinforcement of existing buildings. We perform parametric studies scaling the cross sectional size and reinforcement by the \textit{cross sectional scale factor} $\alpha$ and varying the plastic capacity $\beta$ of the structural elements (see Sec.~\ref{Sims}). In this way, we show the expected collapse mechanisms and the final consequences of progressive collapse in terms of final collapse extent and fragment size distribution for various ($\alpha$, $\beta$). Finally, in Sec.~\ref{RCgen} we compare the damage response of structures with symmetric and asymmetric reinforcement in the beams. 
\section{Simulating Progressive Collapse}\label{Simul}
The choice of spherical DE as simulation tool is motivated by several factors \citep{Poschel_Schwager-2005}: first of all, DE are naturally suitable to deal with dynamic problems since they are based on the direct integration of Newton's equations of motion, which makes the algorithm simple and fast. Moreover, geometric and material nonlinearities, as well as local ruptures can be easily modelled without remeshing. Momentum transmissions due to collisions can be included straightforwardly (see Sec.~\ref{MDmodel}). A quite fine mesh is required to represent the actual volume of the structure and to reduce the error originating from the fact that instead of considering sectional ruptures we instantly remove beam elements that are responsible for the cohesion of the system (see Sec.~\ref{MDmodel}). Considering sectional ruptures would require remeshing while a more precise representation of the volumes could be obtained with polyhedrical DE. For both strategies, the computational demand would grow remarkably. Our model has previously been employed to study fragmentation of materials, e.g.~\citep{Carmona_Wittel-2008}, and its applicability to progressive collapse of structures is demonstrated in \citep{Masoero_Vallini-2010}.

In the simulations, the intact structures are first equilibrated under external service load and gravity. If some elements fail during this initial phase, the structure is incapable of carrying the service load and the simulation is stopped. Differently, if no elements fail, the local damage induced by an accidental event is considered by a sudden removal of a central column of a facade at the first floor (\fignameSP\ref{figMDStruct}), according to the ALPM. The subsequent dynamic stress redistribution can break other elements and trigger widespread progressive collapse. The dynamics of the system is followed by means of explicit time integration, using a 5$^{th}$ order Gear predictor-corrector scheme that, for the explored set of parameters, is stable with time increments lower than $10^{-5}$ seconds. In the following subsections \ref{Struct} and \ref{MDmodel} we give a detailed description of the studied structures and employed model.
\begin{figure}[htb]
\begin{center} \includegraphics[scale=1]{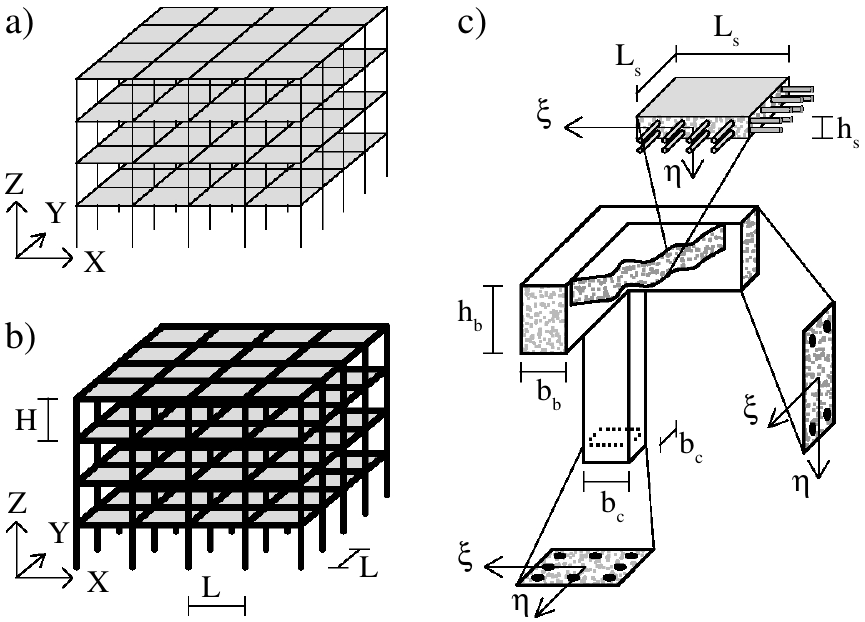}
\caption{The studied framed structure with a) small and b) large $\alpha$. c) Cross sections of the structural elements and arrangement of the reinforcement.}\label{figStruct}
\end{center}
\end{figure}
\subsection{Construction of the framed structures} \label{Struct}
For comparative reasons we limit ourselves to study typical regular 3D frames formed by 4x4x4 identical square cuboid cells with $L$=4m and $H$=3m (see \fignameSP\ref{figStruct}.a). The structures are made of columns along the vertical $Z$ direction, clamped to the ground at $Z = 0$m and connected at each storey by principal beams in $X$ and $Y$ direction. Thin slabs spanning between the principal beams form the floors while the presence of walls is not considered. The geometry of the cross sections of the structural elements is displayed in \fignameSP\ref{figStruct}.c, where the subscripts $c$, $b$ and $s$ denote columns, beams, and slabs. We set the height of the cross sections proportional to the length of the structural element with the coefficients $\lambda$, namely $h_c$=H/10, $h_b$=L/10 and $h_s$=L/50, and we scale each $h$ by a dimensionless \textit{cross sectional scale factor} $\alpha$. $\alpha$ is identical for all elements and enlarges their cross sections making the structure stiffer and stronger. The base edges $b_b$, $b_c$ of the cross section are proportional to the heights $h_b$, $b_c$ with the aspect ratio coefficients $\delta_b$=2/3 and $\delta_c$=1. Consequently, the area, the sectional inertia with respect to the $\xi$ principal direction and the torsional inertia of the cross sections of beams, columns and portions of floor slabs can be easily computed. We represent a structure made of reinforced concrete (RC) with Young's modulus $E_c$ and shear modulus $G_c$ (see Appendix, \tabnameSP\ref{tabMecPar}). The reinforcement is symmetrically distributed (see \fignameSP\ref{figStruct}.c) and its area is proportional to that of the cross section by the factors $\rho_{s,c}$=1.78\%, $\rho_{s,b}$=0.58\%, and $\rho_{s,s}$=1.26\%. The structure carries its own weight $G$, the service external dead load $G_d$=285kg/m$^2$ given by non structural elements like pavement, plaster and internal walls, and the service live load $Q$=667N/m$^2$. $G_d$ and $Q$ are considered uniformly applied to the upper faces of the floors.
\subsection{Model Description}\label{MDmodel}
\begin{figure}[htb]
\begin{center}
\includegraphics[scale=1]{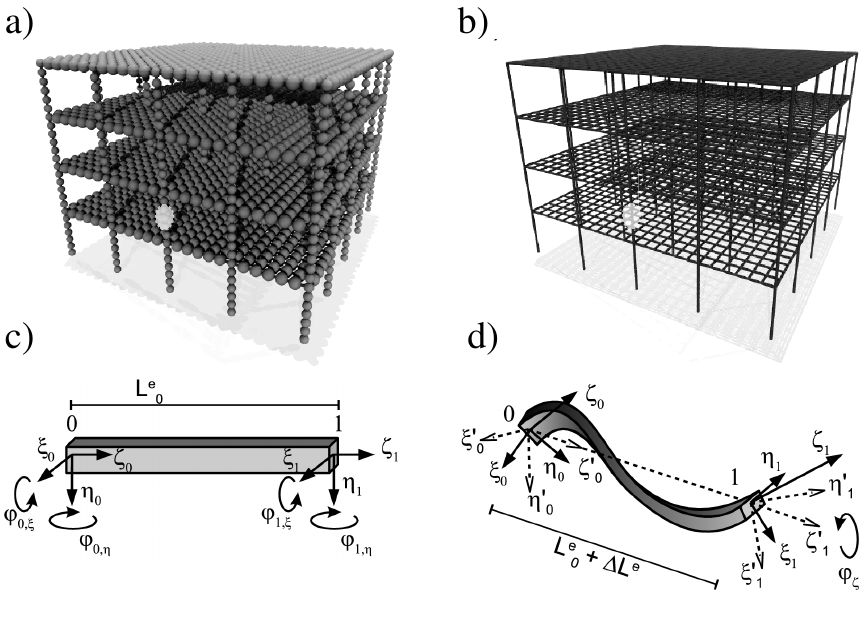}
\caption{a) Spheres and b) EB elements in the modelled structure with $\alpha$ = 1.43. The light grey area marks the initially removed column. Generic Euler-Bernoulli element in c) undeformed and d) deformed state.}\label{figMDStruct}
\end{center}
\end{figure}

We represent the columns and the beams by meshes with $n_c$ and $n_b$ Euler-Bernoulli (EB) beam elements respectively. The floor slabs are considered by a grid of $n_b \times n_b$ EB elements of length $L_s$ (see \fignameSP\ref{figMDStruct}.b) that define slab portions of size $L_s \times L_s \times h_s$ (see \fignameSP\ref{figStruct}.c). To represent the volume of the structure, we set $n_b$ and $n_c$ to obtain $\frac{h^e}{L^e} \; \approx \; 1$, where $h^e$ and $L^e$ are the height of the cross section along $\eta$ and the length of the generic EB element. The cross sections of the EB beam elements are set according to Sec.~\ref{Struct}. The error on $I_{\eta}$ introduced by the simplifying hypothesis $I_{\eta} = I_{\xi}$ is acceptable because the bending of the beams and of the slabs in the horizontal plane is not relevant.

We represent the structural volume by spheres surrounding each node (see \fignameSP\ref{figMDStruct}.a). The diameter of the $k^{th}$ sphere is equal to 90$\%$ of the length of the shortest EB element connected to node $k$. The mass $M_k$ is obtained summing the contributions from the mass $M_{EB}$ of the $m$ EB elements connected to node $k$ and from an extra mass $M_{ex,k}$ given by the external dead load $G_d$: 
\begin{equation} \label{spheremass}
 M_k = \sum_{s=1}^{m} (\frac{1}{2} M_{EB,s}) + M_{ex,k} \;\;.
\end{equation}

Note that in the DE algorithm the EB beam elements do not have a mass since $M_{EB}$ is concentrated in the spheres. The rotational inertia of a sphere is computed considering $M_k$ to be uniformly distributed.
\subsubsection{Euler-Bernoulli beam elements}\label{EB}
The EB elements determine the interactions between pairs of nodes, associating their relative rotations and displacements to forces and moments acting on them. In the following, we describe the linear elastic - perfectly plastic constitutive behavior and the failure rules of the EB elements.
\paragraph{Elastic regime}
In the linear elastic regime, we use the force-displacement relations described in detail in \citep{Carmona_Wittel-2008}, thus taking into account the geometric nonlinearities due to large displacements and neglecting shear deformability. The rotations $\varphi$ are defined starting from the deformed state of the generic EB element (see \fignameSP\ref{figMDStruct}.d). Namely, the bending rotations $\varphi_{\xi}$, $\varphi_{\eta}$ around the $\xi$, $\eta$ principal axes align the $\zeta_0$, $\zeta_1$ axes with the line connecting nodes 0 and 1, while rotating the $\xi'_1$, $\eta'_1$ axes around $\zeta$ by $\varphi_\zeta$ makes them parallel to $\xi'_0$, $\eta'_0$:
\begin{eqnarray}
	\varphi_{0,\xi} , \varphi_{0,\eta} \;    &:&     \; \zeta_0 \rightarrow \zeta'_0 \;\;, \nonumber \\
	\varphi_{1,\xi} , \varphi_{1,\eta} \;    &:&     \; \zeta_1 \rightarrow \zeta'_1 \;\;, \nonumber \\
	\varphi_{\zeta} \;    			 &:&     \; \xi'_1,\eta'_1 \; \parallel \; \xi'_0,\eta'_0 \;\;.
\end{eqnarray}
At a given time step, we compute the axial strain $\varepsilon \; = \; \Delta L^e /L^e_0$ and the rotations $\varphi$ for every EB element starting from the position of the spheres and at the orientation of the $\xi$, $\eta$, $\zeta$ axes frozen to nodes 0 and 1. The forces and moments at nodes 0 and 1 come from the EB beam theory:
\begin{eqnarray} 
	B_{0,j} &=& \frac{E_cI_j^e}{L^e_0}(4\varphi_{0,j}+2\varphi_{1,j}) =		 \frac{E_cI_j^e}{L^e_0}\varphi^{eff}_{0,j}\;\;, 					 \label{M-fi} \\
	B_{1,j} &=& \frac{E_cI_j^e}{L^e_0}(4\varphi_{1,j}+2\varphi_{0,j}) =		 \frac{E_cI_j^e}{L^e_0}\varphi^{eff}_{1,j}\;\;, 					 \label{M1-fi} \\
	T_{i,j} &=& \frac{B_{0,j}+B_{1,j}}{L^e_0}\;\;,                  				\\
	N &=& \pm E_cA^e \varepsilon  \;\;,               \label{N-u}\\
	M_{\zeta} &=& \pm\frac{G_cI_t^e}{L^e_0}\varphi_{\zeta}\;\;. 			\label{Mz-teta} 
\end{eqnarray}
$E_c$ is the Young's modulus of concrete (see \tabnameSP\ref{tabMecPar}). $B_{i,j}$ and $\varphi^{eff}_{i,j}$ denote the bending moment and the effective rotation around the $j=\xi,\eta$ axis at node $i=0,1$, $T_{i,j}$ is the shear force along the $j$ axis at node $i$, $N$ is the normal force and $M_\zeta$ is the torsion. Damping inside the beams is considered adding forces and moments at nodes 0 and 1 that are proportional to the elastic part of the velocities $\dot{\varepsilon}$, $\dot{\varphi_{0,\xi}}$, $\dot{\varphi_{0,\eta}}$, $\dot{\varphi_{1,\xi}}$, $\dot{\varphi_{1,\eta}}$, $\dot{\varphi_{\zeta}}$ by the coefficients in \tabnameSP\ref{tabY-th-damp} but with opposing direction \citep{Poschel_Schwager-2005}.
\paragraph{Plastic regime}
Progressive collapse of structures involves many irreversible processes in the elements and plastic energy dissipation can determine robust or vulnerable responses to damage. To consider plasticity in the EB elements, we make the simplifying assumption that axial and bending plasticization are uncoupled. This choice is justified by our intention to keep the model as simple as possible, leaving refinements to further works. Furthermore, we don't consider plasticization due to shear or torsion in the RC because plastic dissipations associated to them are generally small. Under these hypotheses, the EB elements enter the perfectly plastic regime in axial direction or in bending at node $i$ and around $\xi$ or $\eta$ independently if one or more of these conditions is satisfied:
\begin{eqnarray}
	&& N > N^y \;\; \mbox{or} \;\; N < -|N^y_c|  \;\;,		                  	\label{Ny_cond}\\
	&& |B_{i,j}| > B^y  \;\;. 				\label{By-cond}
\end{eqnarray}
We set the tension and compression yield thresholds $N^y = \rho_s^e A^e f_y$, $N^y_c=A^e f_c$ neglecting the contribution of concrete in tension and steel in compression. The bending yield threshold $B^y$ is evaluated referring to the $\xi$ axis and neglecting the contribution of concrete. If $t_s^e$ is the fraction of reinforcement in tension, i.e. $\frac{3}{8}$ for the columns and $\frac{1}{2}$ for the beams and the slabs (see \fignameSP\ref{figStruct}.c), we obtain: 
\begin{eqnarray} 
	B^{y} = t_s^e \rho_s^e A^e f_y h^e \;\;.		\label{BY}
\end{eqnarray}

We also add a further contribution $\Delta M^y$ to $B^{y}$ to consider that compressions $N<0$ increase $B^{y}$ by reducing the area of concrete in tension during bending. We set $\Delta M^y$ assuming that it is carried by the reinforcement alone and that it compensates the strain $\varepsilon_s$ in the reinforcement in tension produced by $N$, i.e. $\varepsilon_s(N) = \varepsilon_s(\Delta M^y)$:
\begin{equation} 
	\frac{N}{A^eE_c} = \frac{\Delta M^y}{t_s^e\rho_s^eA^eh^eE_s}\;\;.		\label{DMy}
\end{equation}
Within the employed direct time integration scheme, we check yielding in terms of strain and rotations instead of forces and moments. Thus, we adopt elongation $\varepsilon^y$, shortening $\varepsilon^y_c$ and bending $\varphi^{eff,y}$ yield thresholds that satisfy Eqs.~\ref{Ny_cond}~-~\ref{By-cond} in the equality form, when inserted into Eqs.~\ref{M-fi}~-~\ref{Mz-teta}. The expressions for the  $\varepsilon^y$, $\varepsilon^y_c$ and $\varphi^{eff,y}$ are summarized in \tabnameSP\ref{tabY-th-damp}, where also the additional term to $\varphi^{eff,y}$ from Eq.~\ref{DMy} is shown.

\begin{figure}[htb]
\begin{center}
\includegraphics{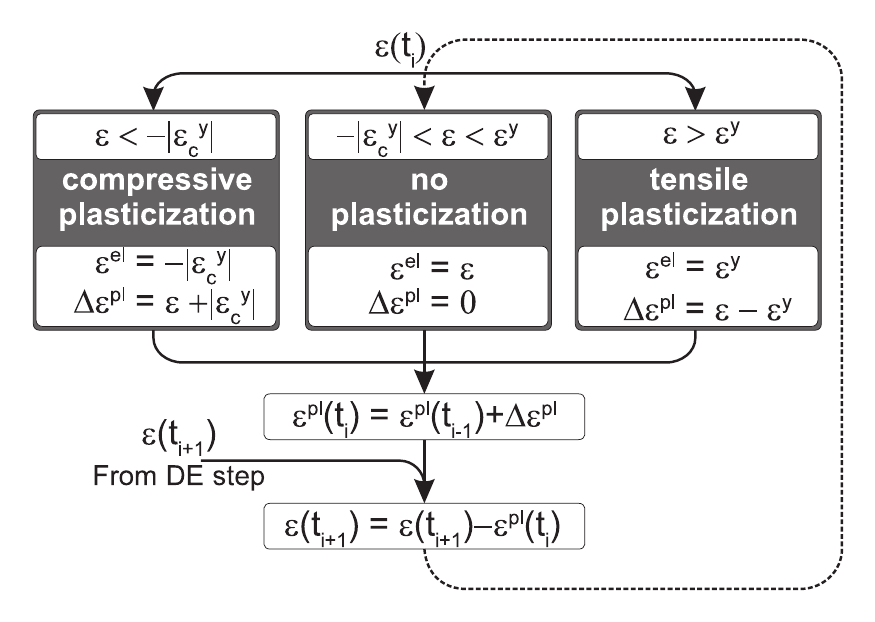}
\caption{Operations to check axial plasticization and to compute the axial plastic flow.} 
\label{figFluxEy}
\end{center}
\end{figure}

In the perfectly plastic regime, we consider the axial strain $\varepsilon$ and rotations $\varphi_0$, $\varphi_1$ to result from the sum of an elastic and a plastic contribution. \fignameSP\ref{figFluxEy} shows how axial plasticization is implemented at the generic time step $t_i$. History dependence is considered accumulating the plastic strain $\varepsilon^{pl}$ in time.

The linear distribution of bending along the EB element makes the description of the rotational plastic regime more complicated. First, we compare $\varphi^{eff}_0$, $\varphi^{eff}_1$ obtained from the integration of Newton's equation with $\varphi^{eff,y}$ to define whether plasticization occurs only at one node or at both. If for instance only node 0 enters the plastic regime, then we set $\varphi_{0}$ = $\varphi^{el}_0$ + $\varphi^{pl}_0$, where $\varphi^{el}_0$, put into Eq.~\ref{M-fi}, satisfies Eq.~\ref{By-cond} in the equality form. Differently, if both node 0 and 1 plasticize, either $\varphi_{0}$ and $\varphi_{1}$ must be separated into elastic and plastic part, so that Eq.~\ref{M-fi} and Eq.~\ref{By-cond} in the equality form return a linear system of two equations in the two unknowns $\varphi^{el}_0$ and $\varphi^{el}_1$. As for axial plasticization, the plastic parts $\varphi^{pl}_0$ and $\varphi^{pl}_1$ must be subtracted from the total $\varphi_{0}$ and $\varphi_{1}$ at the next time step, and the plastic rotations must be cumulated in time.
\paragraph{Element failure} If the strain in a cross section of an EB elements is too high, the element fails and is instantly removed from the system. In the following, we will scale the plastic capacity of the EB elements by a parameter $\beta>$0. If $\beta=0$ the elements break when a combination of the elongation and the effective rotations at nodes $i=0,1$ is large with respect to the yielding thresholds evaluated in uncoupled conditions:
\begin{eqnarray}
\frac{\varepsilon}{\varepsilon^{y}} + max \frac{|\varphi_{i,j}^{eff}|}{\varphi^{eff,y}} \ge 1 	\;\;\; &&\mbox{for}\;\;	\varepsilon>0\;\;, \label{Tbreak_b0}\\	
-\frac{\varepsilon}{|\varepsilon^{y}_c|} + max \frac{|\varphi_{i,j}^{eff}|}{ \varphi^{eff,y}} \ge 1	\;\;\; &&\mbox{for}\;\;	\varepsilon<0 \;\;.      \;\;\;\;\;\;	\label{Cbreak_b0}
\end{eqnarray}

Differently, if $\beta\ge0$, the breaking rules are:
\begin{eqnarray}
\frac{\varepsilon^{pl}}{\beta (\varepsilon^{th}-\varepsilon^{y})} + max \frac{|\varphi_{i,j}^{pl}|}{\beta \varphi^{th}} \ge 1 	\;\;\; &&\mbox{for}\;\;	\varepsilon^{pl}>0 \;\;,  \label{Tbreak}\\	
-\frac{\varepsilon^{pl}}{\beta |\varepsilon^{th}_c-\varepsilon^y_{c}|} + max \frac{|\varphi_{i,j}^{pl}|}{\beta \varphi^{th}} \ge 1	\;\;\; &&\mbox{for}\;\;	\varepsilon^{pl}<0 \;\;,\;\;\;\;\;\;	\label{Cbreak}
\end{eqnarray}
with ultimate threshold values of elongation $\varepsilon^{th}$, shortening $\varepsilon^{th}_c$ and rotation $\varphi^{th}$ estimated in uncoupled conditions. Failure due to shear and torsion is neglected because the shear reinforcement is thought to be designed according to the capacity design approach, that avoids the occurrence of these brittle mechanisms before bending or axial strain failure. We assume $\varepsilon^{th}$ and $\varepsilon^{th}_c$ equal to the ultimate tensile strain of steel and compressive strain of concrete respectively (see \tabnameSP\ref{tabMecPar}). $\varphi^{th}$ is estimated considering a state of uniform bending, and thus uniform curvature $\chi$, in the EB element. Under these hypotheses, the rotation between the edges of an EB element is $\chi L^e$ and gives a strain in the steel bars at $\eta$ = $h^e/2$ equal to $\varepsilon_s=\chi\cdot h^e/2$. We thus obtain $\varphi^{th}$ setting $\varepsilon_s$ equal to the ultimate strain of the steel, i.e. $\varepsilon_s=\varepsilon_{u,s}$. The expressions in \tabnameSP\ref{tabY-th-damp} consider also that the adopted meshing rule assures $h^e/L^e\;\approx\;1$. 
\subsubsection{Hertzian contact between the spheres}\label{MDimpacts}
Progressive failure of EB elements can lead to the free fall of structural elements. We model inter-spheres collisions by a Hertzian potential \citep{Poschel_Schwager-2005,Carmona_Wittel-2008} that generates a conservative repulsive force between partially overlapping spheres. This force is directed along the line connecting the centres of mass of the colliding spheres and is proportional to the overlapping volume by a stiffness parameter $Y$. A similar force is generated when a sphere crosses the $Z = 0$ plane that represents the ground. Impacts dissipates energy due to local fragmentation and to sliding and rolling friction. Thus we introduce forces that are proportional and opposed to the normal, the tangential and the rotational relative velocities of the overlapping spheres with the damping coefficients summarized in \tabnameSP\ref{tabMDimpacts}. In tangential direction, either Coulomb's or dynamic sliding friction are considered.
\section{Parametric studies of Progressive Collapse}\label{Sims}
The plastic capacity of the structural elements is a key factor of progressive collapse. Plasticity determines the subsequent mechanisms of damage propagation based on collisions and the final extent of collapse. We employ the presented model to perform parametric studies on the structural \textit{cross sectional scale factor} $\alpha$ (see Sec.~\ref{Struct}) and on the \textit{plastic capacity} $\beta$ (see Eqs.~\ref{Tbreak} and \ref{Cbreak}). We choose $\alpha$ as parameter because a structure with given $\beta$ can be robust or extremely vulnerable to a column removal, depending on the cross sectional size of the elements. In this way, we can see the effect of plasticity in structures that exhibit different responses to the initial damage, ranging from no collapse to catastrophic collapse. In Sec.~\ref{ColMec} we describe the observed global and local primary mechanisms that can trigger progressive collapse and the subsequent collisions-driven mechanisms. In Sec.~\ref{Collapse} we show the results of the parametric studies. We especially point out what collapse mechanisms occur depending on $\alpha$ and $\beta$ and what are the consequences in terms of final extent of the collapse. Finally, in Sec.~\ref{Frag} we show how $\alpha$ and $\beta$ effect the fragment size distribution of the rubble.
\subsection{Collapse Mechanisms} \label{ColMec}
The initial column removal can trigger three different primary collapse mechanisms that start progressive collapse, two of which are global and lead to total collapse: 

\begin{enumerate}
\item  the first one is caused by elastic waves inside the floors that can separate even distant slabs from the beams (see \fignameSP\ref{figGlobColl}.a).
\begin{figure}[htb]
\begin{center}
\includegraphics[scale=1]{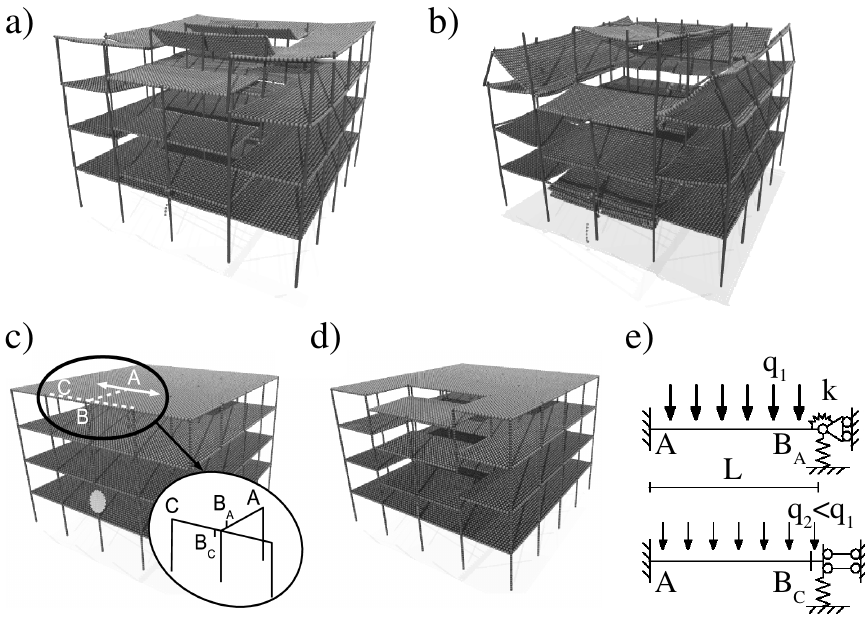}
\caption{Global collapse due to a) catastrophic wave ($\alpha = 0.51$, $\beta = 0.2$) and b) progressive punching ($\alpha = 0.54$, $\beta = 0.8$). Local starting collapse ($\alpha = 0.56$, $\beta = 0.2$): a) detail of the first failure area, where the light gray area marks the initially removed column and the arrows show the direction of crack propagation; b) first stages of the local progressive collapse. e) Approximate static schemes of the elements where the starting damage propagation occurs; the springs in B represent the stiffness of the perpendicular beams.}\label{figGlobColl}
\end{center}
\end{figure}
\item The other global mechanism separates less slabs from the beams but the floors are progressively punched by the columns (see \fignameSP\ref{figGlobColl}.b). Most probably, this mechanism would turn into progressive buckling of the columns if they were less reinforced.
\item The local primary collapse mechanism is characterized by a crack propagating from point A (\fignameSP\ref{figGlobColl}.c) and disconnecting the neighboring floor slabs from the beams (\fignameSP\ref{figGlobColl}.d). 

To explain why rupture occurs at point A instead of point B or C we consider the schematic representation in \fignameSP\ref{figGlobColl}.e. After the column removal, the cross sections $B_A$ and $B_C$, being at two sides of the same node B, show the same vertical displacement. In \fignameSP\ref{figGlobColl}.e, the load on $q_1$ on beam $A-B_A$ is greater than the load $q_2$ on beam $C-B_C$ because the area that can transfer load to the former is larger. Therefore, considering that the torsional stiffness of $C-B_C$ can be represented by a torsional spring that reduces the bending moment in $B_A$ and increases that in $A$, the maximum static bending moment is located in $A$.
\end{enumerate}
\par
\begin{figure}[htb]
\begin{center}
\includegraphics[scale=1]{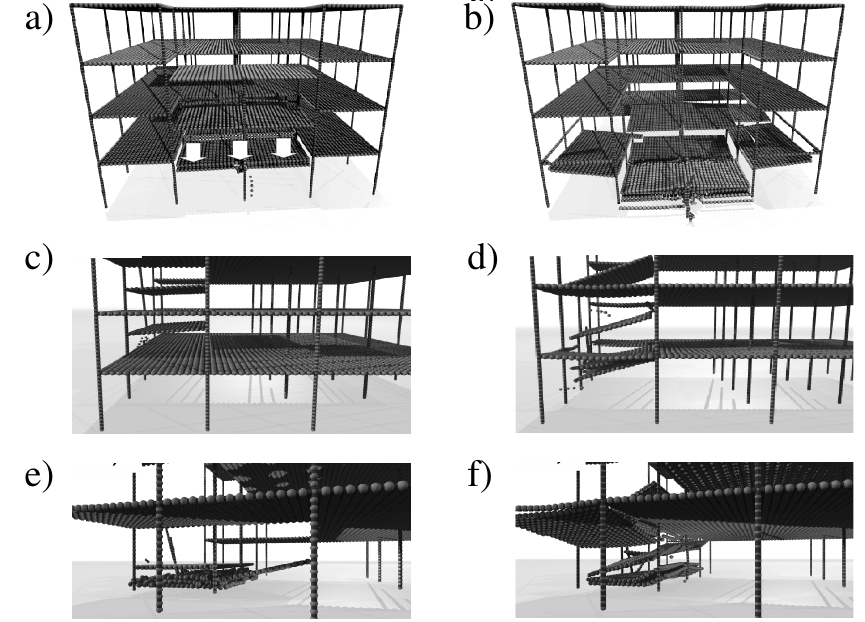}
\caption{a) Hammer and b) subsequent elastic wave at the first storey ($\alpha = 0.54$, $\beta = 0.2$). Free fall of the floor slabs, c) almost horizontal for brittle structures ($\alpha$ = 0.59, $\beta$ = 0.2) and d) tilted for plastic structures ($\alpha = 0.57$, $\beta = 0.8$). In brittle structures e) the slabs stacking on the ground push less against the column than in f) plastic structures}\label{figHamm}
\end{center}
\end{figure}

When a local primary collapse mechanism is triggered, the portion of structure above the removed column undergoes free fall and collides with the floor slabs below. This $hammer$ effect (\fignameSP\ref{figHamm}) generates elastic waves that can damage the neighbouring floor slabs. Rarely does it have catastrophic consequences by itself. 

Other sources of damage transmission are the lateral impacts between falling rubble and still intact portions of the structure. This $drag$ effect can destroy the perimeter beams of neighbouring floor slabs, which can eventually collapse, but usually is not able to cause a widespread propagation of damage by itself. Finally, a severe secondary mechanism is due to the forces exerted by the rubble stacking on the ground, which can cut the neighbouring columns at the base ($base-cutting$, see \fignameSP\ref{figHamm}.c-f).
\subsection{Phase diagram of final states}\label{Collapse}
\begin{figure} [htb]
\begin{center}
\includegraphics[scale=1.00]{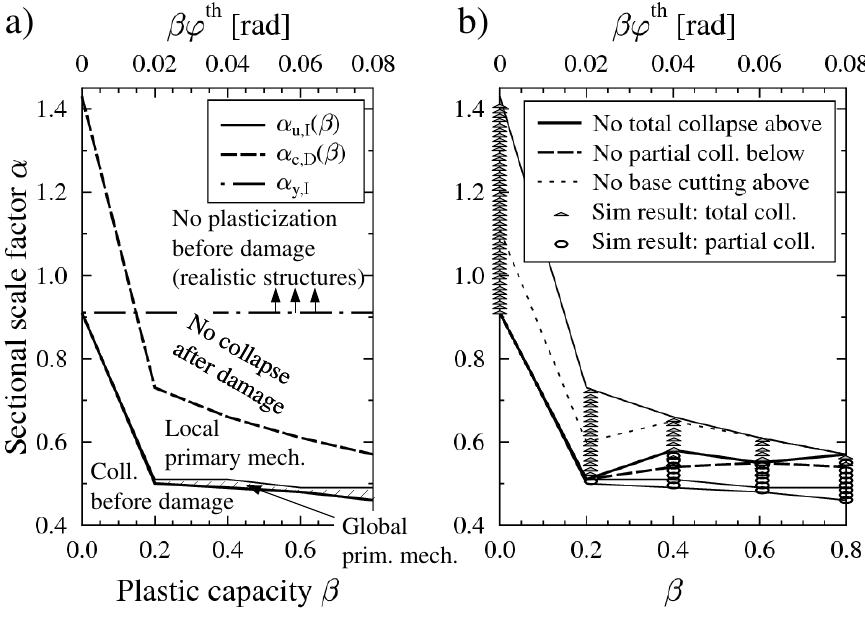}
\caption{Response to the initial damage in terms of final collapse. Above the bold dotted line there is not base cutting. Above the bold dashed line the eventual primary collapse mechanism is local. The markers denote single realizations. $\beta\varphi^{th}$ is the maximum plastic rotation in uncoupled condition (cf. Eqs.~\ref{Tbreak},~\ref{Cbreak})}\label{figPhases}
\end{center}
\end{figure}

The collapse mechanisms described in Sec.~\ref{ColMec} can occur depending on the cross sectional size and reinforcement of the structural elements, determined by the cross sectional scale factor $\alpha$, and on the available plastic capacity $\beta$. \fignameSP\ref{figPhases}.a shows the response of the studied frames to the applied local damage. 

For ($\alpha$, $\beta$) pairs below the $\alpha_{u,I}(\beta)$ curve, the intact structures experience static collapse before the damage. Such weak frames are not supposed to exist since they can not carry the service load. 
Structures with ($\alpha$, $\beta$) within the dashed area just above the $\alpha_{u,I}(\beta)$ curve  completely collapse after the column removal triggers a global primary mechanism. This ($\alpha$, $\beta$) region is narrow and vanishes when $\beta < 0.2$ since brittle failures induce compartmentalisation, abruptly interrupting the dynamic stress flow. 
In this way, either the propagation of waves (dominating when $0.2 < \beta < 0.6$) and the global failure of the storeys due to progressive punching ($\beta > 0.6$) are avoided. Structures with ($\alpha$, $\beta$) above the dashed region and below the $\alpha_{c,D}(\beta)$ exhibit local primary collapse, but can still experience total collapse prompted by the $base-cutting$ mechanism (see \fignameSP\ref{figPhases}.b). If $\beta$ is small, the probability that a sequence of $base-cutting$ provokes total collapse is low, due to the almost null tilt of the floor slabs during the free fall, the high degree of fragmentation after collisions and the larger $\alpha$ values. On the contrary, when $\beta$ is large, sequences of $base-cutting$ are frequent because the slabs tilt while falling and stack on the ground without considerably fragmenting. Consequently, massive slabs lean against the thin (small $\alpha$) base columns at a remarkable height, and cut them (\fignameSP\ref{figHamm}.c-e). 

Partial collapse can occur when $\beta <0.8$ and $\alpha$ is sufficiently large (see \fignameSP\ref{figPhases}.b). In this case all the collision mechanisms of Sec.~\ref{ColMec} determine the final extent of the collapse. Overlapping between the partial and total collapse regions in \fignameSP\ref{figPhases} is due to the fact that the consequences of collision driven mechanism are considerably variable. Structures with ($\alpha,\;\beta$) pairs above the $\alpha_{c,D}(\beta)$ curve are perfectly robust, i.e. they do not suffer any further failure after the column removal. Finally, when $\alpha >\alpha_{y,I}$ the intact structure does not plasticize in the pre-dmage stage, i.e. during the static application of the service load. Since this is a common requirement for buildings in service conditions, structures with realistic size of the elements are located in the $\alpha > \alpha_{y,I}$ region.
\begin{figure}[htb]
\begin{center}
\includegraphics[scale=1.0]{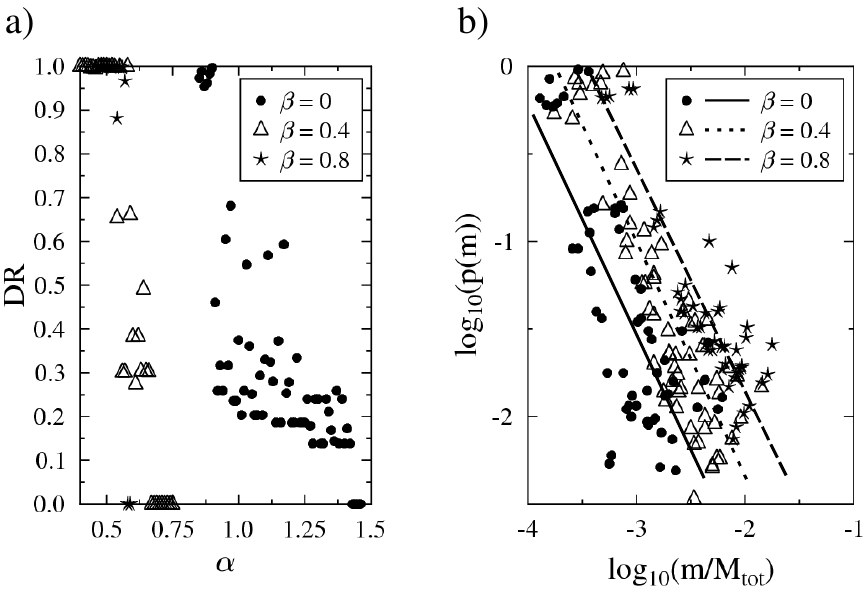}
\caption{a) Final entity of the collapse for ($\alpha$, $\beta$) pairs between $\alpha_{u,I}(\beta)$ and $\alpha_{c,D}(\beta)$. Note that for each $\beta$, values of $\alpha$ larger or smaller than those of the plotted points give respectively $DR = 0$ and $DR = 1$. b Fragment mass distribution for different plastic capacities $\beta$. $p(m)$ is the fraction of fragments with mass $m$ normalised to the total mass of the structure $M_{tot}$.}\label{figCol-Alp}
\end{center}
\end{figure}

The beneficial effect of plasticity is evident from the fact that $\alpha_{c,D}(\beta)$ and $\alpha_{u,I}(\beta)$ decrease with $\beta$ (see \fignameSP\ref{figPhases}.a). These curves sensibly decrease in the range  $0<\beta<0.2$, which means that the provision of a minimal plastic capacity is sufficient to remarkably improve the structural response, assuring complete safety to structures with $\alpha  >  \alpha_{y,I}$. This result is a consequence of the symmetric distribution of reinforcement inside the beams and the floor slabs, that makes their bending behavior qualitatively similar to that of steel elements. Steel structures are actually likely to sustain one column removal without damage propagation. In Sec.~\ref{RCgen} we show that RC structures with realistic cross sectional size of the elements and  asymmetric reinforcement distribution would experience partial collapse. Differently, when $\beta>0.4$, $\alpha_{c,D}(\beta)$ and $\alpha_{c,D}(\beta)$ do not decrease much anymore (see \fignameSP\ref{figPhases}.a). This means that the initiation of progressive collapse is a local phenomenon associated with relatively small plastic stress redistributions, as already observed for 2D steel frames by \citep{Khandelwal_El-Tawil-2008}.

\fignameSP\ref{figCol-Alp}.a shows a quantitative measure of the final collapse extent for structures with different plastic capacity $\beta$. In particular, we compute the \textit{demolition ratio} $DR$, i.e. the fraction of lost living space at the end of the collapse. From the figure it can be immediately seen that structures with high plastic capacity $\beta$ undergo progressive collapse only if they have thin elements, i.e. small $\alpha$, but if the collapse is triggered it will affect the whole system. The large variability of $DR$ is due to the collision-driven mechanisms and, in particular, the largest values of $DR$ associated with partial collapse, i.e. when $DR\ne0$ and $DR\ne1$, indicate the occurrence of $base-cuttings$.
\subsection{Fragment Size Distribution}\label{Frag}
The size of the fragments produced after a structural collapse is interesting for controlled demolitions, where large fragments require further effort to relocate them. 
The probability density distribution $p$ of the mass of the fragments $m$ normalised by the total mass of the structure $M_{tot}$ is shown in \fignameSP\ref{figCol-Alp}.b. 
We observed that $\alpha$ does not influence $p(m)$. Differently, as the plastic capacity of the elements $\beta$ grows the power law regression lines in figure shift towards larger sizes of the fragments while their exponent, between -1.2 and -1.4, does not seem to change according to a trend. Note that this exponent is in the range of the one reported for shell fragmentation of 1.35 \citep{Wittel-2004}. When $\beta$ is small $p(m)$ has large dispersion and most of the fragments are represented by single spheres completely disconnected from the others. Calling $M_1$ the sum of the masses of the fragments made of one sphere and  $M_{F}$ the total mass of the fragments, $M_1/M_{F}$ is larger than 0.7 when $\beta=0$. This denotes a finite size effect, i.e. the single sphere is larger than the characteristic size of the fragments. On the contrary, when $\beta$ is high $p(m)$ is less dispersed and $M_1/M_{F}< 0.25$ implies that the fragment size distribution is better caught.
\begin{figure}[htb]
\begin{center}
\includegraphics[scale=1.0]{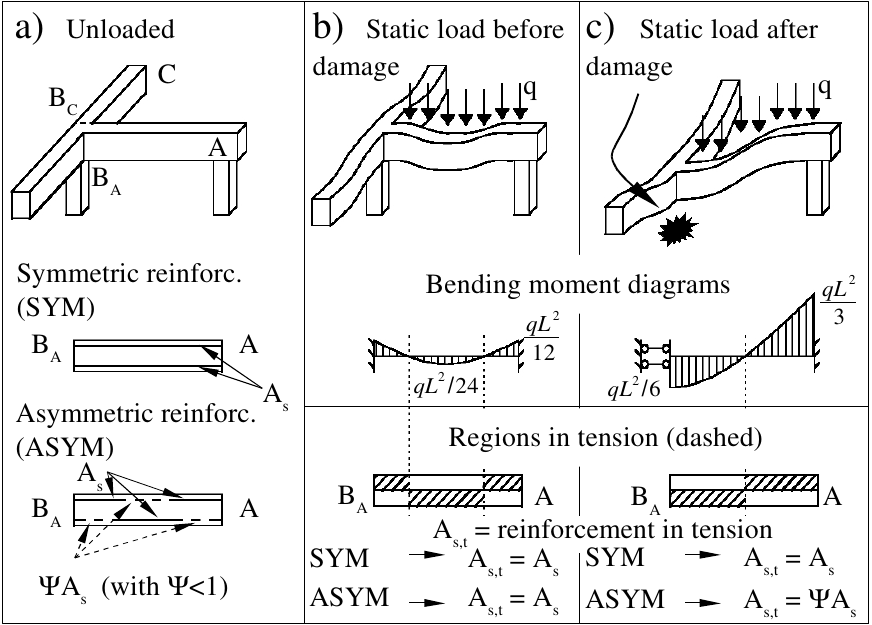}
\caption{3D representation of the beams involved in the local primary collapse mechanism in Fig. \ref{figGlobColl}.c,d) Symmetric and asymmetric reinforcement arrangement. Static deformed condition of the beams, bending moment in $A-B_A$ and regions in tension b) before and c) after the column removal.}\label{figAsym}
\end{center}
\end{figure}
\section{The effect of asymmetric reinforcement}\label{RCgen}
The results in Sec.~\ref{Collapse} refer to RC structures with symmetrically reinforced elements. Nevertheless, the reinforcement inside the beams and the floor slabs of real structures is mostly concentrated in the regions under tension in service conditions. 
\fignameSP\ref{figAsym}.a shows the symmetric and the more realistic asymmetric reinforcement arrangement inside the beams that are involved in the local primary collapse mechanism of \fignameSP\ref{figGlobColl}.c,d. 
The column loss in \fignameSP\ref{figAsym}.c produces the inversion of the bending moment in $B_A$ and thus, in case of asymmetric reinforcement, the reinforcement under tension $A_{s,t}$ in $B_A$ passes from $A_s$ before the damage (see \fignameSP\ref{figAsym}.b) to $\Psi A_s$ after the damage (see \fignameSP\ref{figAsym}.c). Therefore, if $\Psi$ is small, section $B_A$ is likely to fail before section $A$, differently from the observations in case of symmetric reinforcement (see Sec.~\ref{ColMec}).
In the following, depending on $\Psi$, we estimate the $\alpha_{c,D}(\beta)$ curves that separates the collapse and robustness regions in \fignameSP\ref{figPhases}.a for frames with asymmetric reinforcement.
\begin{figure}[htb]
\begin{center}
\includegraphics[scale=1.0]{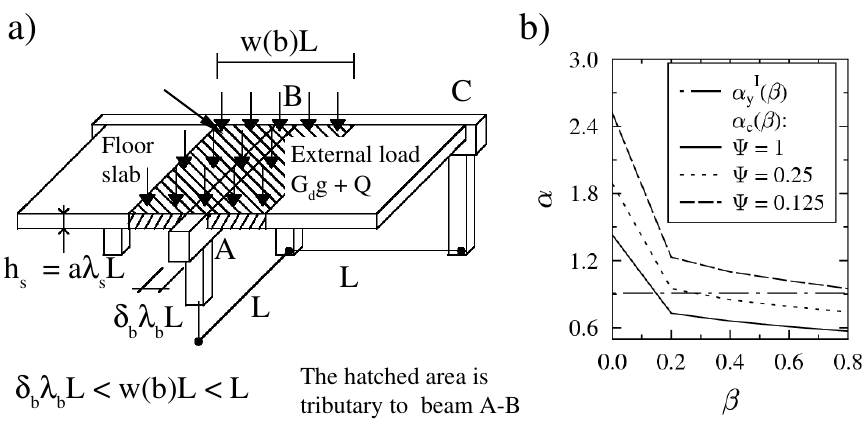}
\caption{a) Load scheme ofthe the $A-B$ beam (cf. Eq.~\ref{q(alfa)}). b) Transition to no collapse after the column removal for different ratios of asymmetry $\psi$ of the reinforcement in the beams.}\label{figLoad}
\end{center}
\end{figure}

Failure in $B_A$ occurs if:
\begin{equation} \label{BBA=BYBA}
	B_{B_A} = \frac{qL^2}{6} = DYN(\beta) \cdot B^y_{B_A} \;\;,
\end{equation}
where $B_{B_A}$ is the static bending moment in $B_A$ after the damage (see \fignameSP\ref{figAsym}.c) and $DYN(\beta)$ is an amplifying factor that considers dynamics and plastic capacity. $DYN$ ranges from $\frac{1}{2}$ for linear elastic - perfectly brittle structures ($\beta=0$) to $1$ for perfectly plastic structures with $\beta=\infty$. $q$ is the load per unit length on the beam (see \fignameSP\ref{figLoad}.a): 
\begin{equation} \label{q(alfa)}
	q = [Q + (G_d+\lambda_s \alpha L \gamma_c)g]\varpi(\beta) L \;\;,
\end{equation}
$B^y_{B_A} = \Psi B^y$ is the yield bending moment in $B_A$, with $B_y\propto\alpha^3$ because it is proportional to the area of the cross section times its height (see Eq.~\ref{BY}). 
Thus, Eq.~\ref{BBA=BYBA} can be rewritten as.
\begin{equation} \label{alfac}
	[Q + (G_d+\lambda_s \alpha L \gamma_c)g]\varpi(\beta) \frac{L^3}{6} = DYN\cdot\Psi t_s\rho_{s,b}\lambda_b^3\delta_b f_y \alpha^3 \;\;.
\end{equation}
For a given $\Psi$, solving Eq.~\ref{alfac} in $\alpha$ for different $\beta$ permits to trace the $\alpha_{c_D}(\beta)$ curves. Nevertheless, in Eq.~\ref{alfac} the parameter $DYN/\varpi(\beta)$ is still unknown. 

In order to assign values to $DYN/\varpi(\beta)$, we repeat the previous argument for structures with symmetric reinforcement, for which we now the $\alpha_{c_D}(\beta)$ values at some $\beta$ obtained from our simulations (see \fignameSP\ref{figPhases}.a). In case of symmetric reinforcement, the $A-B_A$ beam in \fignameSP\ref{figAsym}.c fails in $A$, where the bending moment in $qL^2/3$. Since in this case $\Psi=1$, Eq. \ref{alfac} turns into:
\begin{equation} \label{alfac_sym}
	[Q + (G_d+\lambda_s \alpha L \gamma_c)g]\varpi(\beta) \frac{L^3}{3} = DYN\cdot t_s\rho_{s,b}\lambda_b^3\delta_b f_y \alpha^3 \;\;.
\end{equation}
Solving Eq.~\ref{alfac_sym} starting from the $\alpha_{c_D}(\beta)$ values in \fignameSP\ref{figPhases}.a, we obtain the $DYN/\varpi(\beta)$ values in \tabnameSP\ref{DYN_omega}.
\renewcommand{\arraystretch}{1.5}

\begin{table}
\begin{center}
\caption{$DYN/\varpi$ factors of Eqs.~\ref{alfac} and \ref{alfac_sym}.}\label{DYN_omega} 
{
\begin{tabular}{|p{1.5cm}|c|c|c|c|c|}
\hline
$\beta$   	&    0.0    &  0.2	& 0.4	& 0.6	& 0.8	\\
\hline
$DYN/\varpi$   	&    0.85    &  5.00	& 6.60	& 8.20	& 9.80	\\
\hline
\end{tabular}
}
\end{center}
\end{table}
\renewcommand{\arraystretch}{1.0}

Inserting the $DYN/\varpi(\beta)$ values from \tabnameSP\ref{DYN_omega} into Eq.~\ref{alfac}, we obtain the curves in \fignameSP\ref{figLoad}.b. These curves show that collapse can initiate also in well designed RC structures, i.e. with $\alpha>\alpha_{y,I}$. Moreover, we argue that the final extent of the collapse should be partial. Note that the $\alpha_{u.I}(\beta)$ curve in \fignameSP\ref{figPhases} does not change with $\psi$ since it only depends on the reinforcement in tension in the intact structure. Also, the bold dotted line in \fignameSP\ref{figPhases}.b regarding the base-cutting phenomenon does not depend on $\psi$ but only on the reinforcement inside the columns, that is usually symmetric. Therefore, when $\psi<1$ an expansion of the partial collapse region towards higher values of $\alpha$ is expected.
\section{Conclusions and Outlook}\label{Concl}
Progressive collapse of framed structures after local damage consist in an initial triggering and a subsequent damage propagation. If the initial damage is small, like the studied column removal, collapse initiation is generally a local phenomenon affecting the surroundings of the initially damaged area. Global primary mechanisms can occur only in thin structures with enough plastic capacity to avoid the compartmentalisation effect produced by brittle ruptures, i.e. $\beta>0.2$. Nevertheless, if the starting damage is more serious than a single column removal, global primary mechanisms can be expected also for larger and more brittle structures. In case of multiple column removal, progressive crushing or buckling of the columns is a possible global primary collapse mechanism that was not observed in this context. We showed that structures with minimal plastic capacity and symmetrically reinforced beams are robust towards a single column removal. On the contrary, frames with asymmetric reinforcement would experience partial collapse even if made of elements with large plastic capacity.

If a local primary collapse mechanism is triggered, the final extent of the collapse depends on secondary mechanisms driven by collisions between the structural elements. We showed that damage can not widely propagate in structures with small plastic capacity since brittle failures compartmentalise the system. Differently, structures with large plastic capacity tend to collapse entirely after a sequence of base cutting. This result is a consequence of the fact that, in presence of large plastic capacity, we studied frames with thin structural elements. In facts, columns with larger and thus more realistic cross sectional size and reinforcement would not fail because of base cutting.

Collision-driven mechanisms also determine the outcome of the fragmentation process. We showed that the fragment mass distribution does not depend on the strength and stiffness of the structural elements. It is namely influenced by the plastic capacity of the elements. In structures with large plastic capacity the fragments are more massive and represent an extra cost in controlled demolitions processes.

In the present paper, for clearer interpretability of the results, we limited ourselves to simple geometry, collisions, and constitutive models. Implementing more sophisticated collision models, e.g. using polyhedral discrete elements, or enabling discrete elements to fragment (see e.g.~\citep{Poschel_Schwager-2005}), as well as rate effects are future challenges. We also neglected shear failures, since the structural elements were sufficiently small and slender, but this hypothesis should be removed to deal with structures made of large elements. These models should also be refined if the aim is to simulate in detail the collapse of specific real buildings. Nevertheless, already at the present state, several interesting studies can be conducted to analyze the response to earthquakes and to investigate the influence of material disorder, geometric uncertainties, overall geometry, and structural connections. Experimental validation remains problematic in the field of progressive collapse, because of difficulty in monitoring collapse of complex buildings, and due to problems concerning repeatability of experiments. Further discussion about monitoring collapse of buildings for model validation, especially concerning demolitions, can be found in \citep{Bazant_Verdure-2007}.


\pagebreak
\newpage

\appendix
\section{Tables of parameters}
\begin{table}[htp]
	\begin{center}
	\caption{Loads and mechanical properties of the materials and of the frames.} \label{tabMecPar}
{
\begin{tabular}{|p{4.1cm}|>{\centering\arraybackslash}p{1.2cm}|>{\centering\arraybackslash}p{.8cm}|>{\centering\arraybackslash}p{1.3cm}|}
\hline
$Parameter$   			&    $Symbol$    	&   $Units$ 		& $Value$ 	\\
\hline
\hline
\multicolumn{4}{|l|}{$Properties \; of \;the \; concrete$} \\
\hline
Specific weight			&   $\gamma_{RC}$	&  kg/m$^3$		& 2500\\
Young modulus			&   $E_c$     		&  N/m$^2$   		& 30$\cdot 10^9$ \\
Shear modulus			&   $G_c$		&  N/m$^2$		& $15\cdot 10^9$ \\
Compressive yield stress	&   $f_c$		&  N/m$^2$		& 20$\cdot 10^6$\\
Ultimate shortening 		&   $\epsilon_{u,c}$	&  -			& 0.0035	\\
\hline
\hline
\multicolumn{4}{|l|}{$Properties \; of \;the \; steel$} \\
\hline
Young's modulus			&   $E_s$		&  N/m$^2$		& 200$\cdot 10^9$ \\
Yield stress			&   $f_y$		&  N/m$^2$		& 440$\cdot 10^6$\\	
Yield strain			&   $\epsilon_{y,s}$	&  -			& 0.0022 \\
Ultimate strain			&   $\epsilon_{u,s}$	&  -			& 0.05 \\
\hline
\end{tabular}
}
	\end{center}
\end{table}

\begin{table}[htp]
	\begin{center}
	\caption{Damping, yielding and failure parameters of the Euler-Bernoulli elements.} \label{tabY-th-damp} 
{
\begin{tabular}{|p{2.4cm}|>{\centering\arraybackslash}p{1.2cm}|>{\centering\arraybackslash}p{0.8cm}|>{\centering\arraybackslash}p{3.0cm}|}
\hline
$Parameter$   			&    $Symbol$    	&   $Units$ 		& $Value$ 	\\
\hline
\hline
\multicolumn{4}{|l|}{$Damping \; coefficients$} \\
\hline
Elongation			&   $\gamma_L$		&  Ns/m  	& $100$   \\
Torsion 			&   $\gamma_T$		&  Nms	& $1$   \\
Bending 		&   $\gamma_B$		&  Nms	& $10$   \\
\hline
\hline
\multicolumn{4}{|l|}{$Axial \; yielding$} \\
\hline
Elongation			& $\varepsilon^{y}$	&  -			& $N^y/(E_cA^e)$ \\
Compression			& $\varepsilon^y_c$	&  -			& $N^y_{c}/(E_cA^e)$ \\
\hline
\hline
\multicolumn{4}{|l|}{$Effective \; rotation \; yielding$} \\
\hline
Columns 			& $\varphi^{eff,y}$	&  rad		& $B^yL^e/(E_cI^e_{\xi})+$\\
				&			&			& $ +\varepsilon9L^e\rho_sE_s/(2h^eE_c)$\\
Beams \& slabs	& $\varphi^{eff,y}$	&  rad		& $B^yL^e/(E_cI^e_{\xi})+$\\
				&			&			& $   +\varepsilon6L^e\rho_sE_s/(h^eE_c)$\\
\hline
\hline
\multicolumn{4}{|l|}{$Ultimate \; thresholds$} \\
\hline
Elongation			& $\varepsilon^{th}$	&  $-$			& $\varepsilon_{u,s}$	\\
Compression 			& $\varepsilon^{th}_c$	&  $-$			& $\varepsilon_{u,c}$	\\
Rotation 			& $\varphi^{th}$	&  $rad$		& $2\varepsilon_{u,s}L^e/h^e \approx $\\
				&			&			& $\approx 2\varepsilon_{u,s} = 0.1$\\
\hline
\end{tabular}
}
	\end{center}
\end{table}

\begin{table}[htp]
	\begin{center}
	\caption{Contact parameters.} \label{tabMDimpacts} 
{
\begin{tabular}{|p{3.6cm}|>{\centering\arraybackslash}p{1.2cm}|>{\centering\arraybackslash}p{0.8cm}|>{\centering\arraybackslash}p{1.6cm}|}
\hline
$Parameter$   			&    $Symbol$    	&   $Units$ 		& $Value$ 	\\
\hline
\hline
\multicolumn{4}{|l|}{$Overlap \; stiffness$} \\
\hline
Sphere-sphere			&   $Y$			&  N/m$^3$	& $10^7$ \\
Sphere-ground			&   $Y^g$			& N/m$^3$	& $5\cdot 10^7$ \\
\hline
\hline
\multicolumn{4}{|l|}{$Normal \; damping \; coefficients$} \\
\hline
Sphere-sphere			&   $\gamma_n$ 		&  Ns/m	& $5\cdot 10^4$ \\
Sphere-ground			&   $\gamma_n^g$ 	&  Ns/m	& $10^5$ \\
\hline
\hline
\multicolumn{4}{|l|}{$Sphere-sphere \; tangential \; damping \; coefficients$} \\
\hline
Coulomb friction		&  $\mu$		&  Ns/m	& $5\cdot 10^3$ \\
Dynamic friction		&  $\gamma_t$		&  Ns/m	& $5\cdot 10^3$ \\
Rolling	friction		&   $\gamma_w$		&  Nms	& $5\cdot 10^1$	\\
\hline
\hline
\multicolumn{4}{|l|}{$Sphere-ground \; tangential \; damping \; coefficients$} \\
\hline		
Coulomb friction		&  $\mu^g$		&  Ns/m	& $10^4$ \\
Dynamic friction		&  $\gamma_t^g$		&  Ns/m	& $10^4$ \\
Rolling friction		&   $\gamma_w^g$	&  Nms	& $10^2$	\\
\hline
\end{tabular}
}
	\end{center}
\end{table}

\end{document}